\documentclass{article}

     \usepackage[nonatbib,preprint]{neurips_2020}

\usepackage[utf8]{inputenc} 
\usepackage[T1]{fontenc}    
\usepackage{hyperref}       
\usepackage{url}            
\usepackage{booktabs}       
\usepackage{amsfonts}       
\usepackage{nicefrac}       
\usepackage{microtype}      
\usepackage{graphicx}       
\usepackage{subcaption}     
\usepackage{adjustbox}

\usepackage[citestyle=numeric-comp, sorting=none, giveninits=true]{biblatex}
\title{COVID-Net CXR-2: An Enhanced Deep Convolutional Neural Network Design for Detection of COVID-19 Cases from Chest X-ray Images}

\author{Maya Pavlova$^{1}$, Naomi Terhljan$^{1}$, Audrey G. Chung$^{2,3}$, Andy Zhao$^{1}$\\ \textbf{Siddharth Surana$^{1}$, Hossein Aboutalebi$^{1,2}$, Hayden Gunraj$^{1}$}\\
\textbf{Ali Sabri$^{4}$, Amer Alaref$^{5,6}$, and Alexander Wong$^{1,2,3}$}\\
$^1$University of Waterloo, Canada\\
$^2$Waterloo Artificial Intelligence Institute, Canada\\
$^3$DarwinAI Corp., Canada\\
$^4$Department of Radiology, Niagara Health, McMaster University, Canada\\
$^5$Department of Diagnostic Radiology, Thunder Bay Regional Health Sciences Centre, Canada\\
$^6$Department of Diagnostic Imaging, Northern Ontario School of Medicine, Canada\\
}

\begin{document}

\maketitle

\begin{abstract}
   As the COVID-19 pandemic continues to devastate globally, the use of chest X-ray (CXR) imaging as a complimentary screening strategy to RT-PCR testing continues to grow given its routine clinical use for respiratory complaint. As part of the COVID-Net open source initiative, we introduce COVID-Net CXR-2, an enhanced deep convolutional neural network design for COVID-19 detection from CXR images built using a greater quantity and diversity of patients than the original COVID-Net. To facilitate this, we also introduce a new benchmark dataset composed of 19,203 CXR images from a multinational cohort of 16,656 patients from at least 51 countries, making it the largest, most diverse COVID-19 CXR dataset in open access form. The COVID-Net CXR-2 network achieves sensitivity and positive predictive value of 95.5\%/97.0\%, respectively, and was audited in a transparent and responsible manner. Explainability-driven performance validation was used during auditing to gain deeper insights in its decision-making behaviour and to ensure clinically relevant factors are leveraged for improving trust in its usage. Radiologist validation was also conducted, where select cases were reviewed and reported on by two board-certified radiologists with over 10 and 19 years of experience, respectively, and showed that the critical factors leveraged by COVID-Net CXR-2 are consistent with radiologist interpretations.  While not a production-ready solution, we hope the open-source, open-access release of COVID-Net CXR-2 and the respective CXR benchmark dataset\footnote{\url{http://www.covid-net.ml}} will encourage researchers, clinical scientists, and citizen scientists to accelerate advancements and innovations in the fight against the pandemic.

\end{abstract}

\section{Introduction}

As the global devastation of the coronavirus disease 2019 (COVID-19) pandemic continues, the need for effective screening methods has grown. A crucial step in the containment of the severe acute respiratory syndrome
coronavirus 2 (SARS-CoV-2) virus causing the COVID-19 pandemic is effective screening of patients in order to provide immediate treatment, care, and isolation precautions. While the main screening method is reverse transcription polymerase chain reaction (RT-PCR) testing~\cite{Wang2020_RTPCR}, recent studies have shown that the sensitivity of such tests can be relatively low and highly variable depending on how or when the specimen was collected~\cite{Fang2020,Yang2020, Li2020_RTPCR, Ai2020}. 

Chest X-ray (CXR) radiography screenings are a complimentary screening method to RT-PCR that has seen growing interest and increased usage in clinical institutes around the world. Studies have shown characteristic pulmonary abnormalities in SARS-CoV-2 positive cases such as ground-glass opacities, bilateral abnormalities, and interstitial abnormalities~\cite{Wong,Warren,Toussie,Huang,Guan}. Compared to other imaging modalities, CXR equipment is readily available in many healthcare facilities, is relatively easy to decontaminate, and can be used in isolation rooms to reduce transmission risk~\cite{RSNA} due to the availability of portable CXR imaging systems~\cite{Jacobi}. More importantly, CXR imaging is a routine clinical procedure for respiratory complaint~\cite{BSTI}, and thus is frequently conducted in parallel with viral testing to reduce patient volume.  

Despite the growing interest and usage of CXR radiography screenings in the COVID-19 clinical workflow, a challenge faced by clinicians and radiologists during CXR screenings is differentiating between SARS-CoV-2 positive and negative infections. More specifically, it has been found that potential indicators for SARS-CoV-2 infections may also be present in non-SARS-CoV-2 infections, and the differences in how they present can also be quite subtle.  As such, computer-aided screening systems are highly desired for assisting front-line healthcare workers to streamline the COVID-19 clinical workflow by more rapidly and accurately interpreting CXR images to screen for COVID-19 cases.

Motivated by this, we launched the COVID-Net open source initiative\footnote{\url{http://www.covid-net.ml}}~\cite{covidnet,alex2020covidnets,Gunraj2020,gunraj2021covidnet,ebadi2021covidxus,aboutalebi2021covidnet,wong2021tbnet,zhang2021covid19,aboutalebi2021covidnet2} for accelerating the advancement and adoption of deep learning for tackling this pandemic. 
While the initiative has been successful and leveraged globally, the continuously evolving nature of the pandemic and the increasing quantity of available CXR data from multinational cohorts has led to a growing demand for ever-improving computer-aided diagnostic solutions as part of the initiative.  Since the launch of the COVID-Net open source initiative, there have been many studies in the area of COVID-19 case detection using CXR images, with many leveraging the open access datasets and open source deep neural networks made publicly available through this initiative~\cite{li2020covidmobilexpert, minaee2020deepcovid, afshar2020covidcaps, luz2020effective, Khobahi2020.04.14.20065722,ucar,tartaglione2020unveiling,yeh2020cascaded,zhang2020covidda,karim2020deepcovidexplainer,Apostolopoulos_2020,farooq2020covidresnet,Majeed2020,MONSHI2021104375,JIA2021104425,Ismael,Abbas}.

In this work, we introduce COVID-Net CXR-2, an enhanced deep convolutional neural network design for COVID-19 chest X-ray detection built on a greater quantity and diversity of patients than the original COVID-Net network design~\cite{covidnet}. To facilitate this, we introduce a benchmark dataset that is, to the best of the authors' knowledge, the largest, most diverse open access COVID-19 CXR cohort, with patients from at least 51 countries. We leverage explainability-driven performance validation to audit COVID-Net CXR-2 in a transparent and responsible manner to ensure the decision-making behaviour is based on relevant visual indicators for improving trust in its usage.  Furthermore, radiologist validation was conducted where select cases were reviewed and reported on by two board-certified radiologists with over 10 and 19 years of experience, respectively.  While not a production-ready solution, we hope the open-source, open-access release of COVID-Net CXR-2 and the respective CXR benchmark dataset will help encourage researchers, clinical scientists, and citizen scientists to accelerate advancements and innovations in the fight against the pandemic.

\begin{figure*}[!t]
  \centering
  \includegraphics[width= \linewidth]{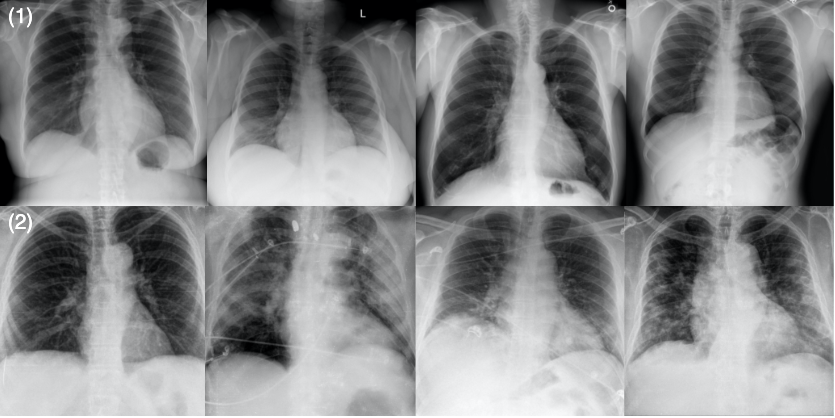}
  \caption{Example chest X-ray images from the benchmark dataset: (1) SARS-CoV-2 negative patient cases and (2) SARS-CoV-2 positive patient cases.}
  \label{fig:cxr-examples}
\end{figure*}

The paper is organized as follows. Section~\ref{sec:methods} describes the underlying methodology behind the construction of the proposed COVID-Net CXR-2 as well as the preparation of the benchmark dataset.  Section~\ref{sec:results} presents and discusses the efficacy and decision-making behaviour of COVID-Net CXR-2 from both a quantitative perspective as well as a qualitative perspective.  Finally, conclusions are drawn in Section~\ref{sec:conclusion}.

\section{Methods}\label{sec:methods}

In this study, we introduce COVID-Net CXR-2, an enhanced deep convolutional neural network design for detection of COVID-19 from chest X-ray images. To train and test the network, we further introduce a new CXR benchmark dataset which represents the largest, most diverse open access COVID-19 CXR dataset available, spanning a multinational patient cohort from at least 51 countries.  The details regarding data preparation, network design, and explainability-driven performance validation are described below.

\subsection{Benchmark dataset preparation}
\begin{figure*}[!t]
  \centering
  \includegraphics[width=0.7\linewidth]{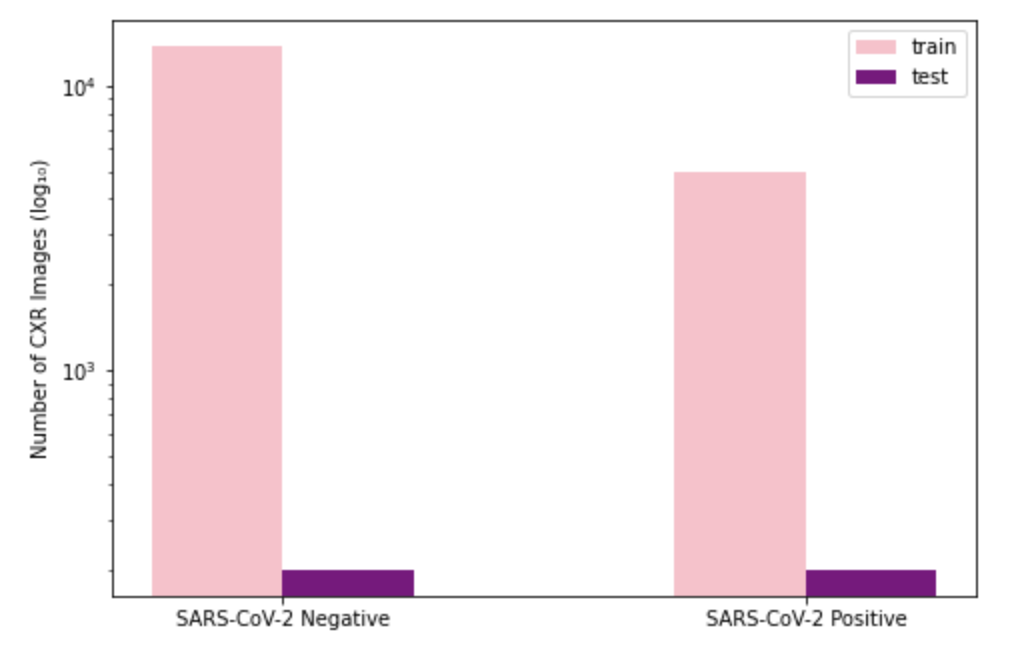}
  \caption{Image-level distribution of benchmark dataset for SARS-CoV-2 negative and positive cases. (Left bar) number of training images, (right bar) number of test images.}
  \label{fig:image-dist}
\end{figure*}

\begin{figure*}[!t]
  \centering
  \includegraphics[width=0.7\linewidth]{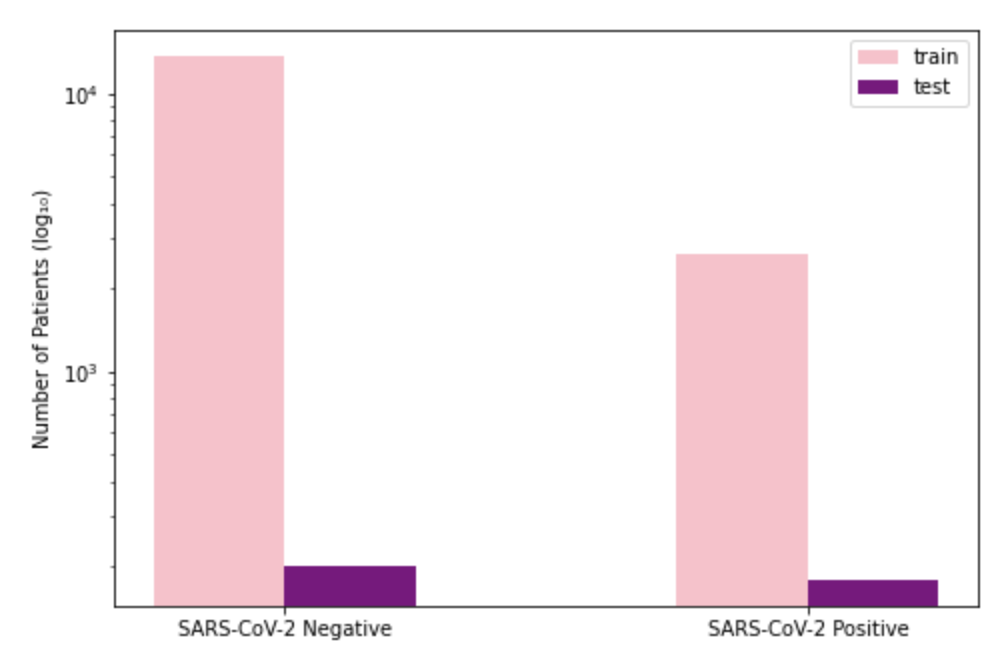}
  \caption{Patient distribution of benchmark dataset for SARS-CoV-2 negative and positive cases. (Left bar) number of training patients, (right bar) number of test patients.}
  \label{fig:patient-dist}
\end{figure*}

To train and evaluate COVID-Net CXR-2, we first created a new CXR benchmark dataset with example images shown in Figure~\ref{fig:cxr-examples}, unifying patient cohorts from several organizations and initiatives from around the world~\cite{kaggle,ricord,kaggle2,Figure1,vaya2020bimcv,actualmed,Cohen}. The new CXR benchmark dataset comprises 19,203 CXR images from a multinational cohort of 16,656 patients from at least 51 countries, making it the largest, most diverse COVID-19 CXR dataset available in open access form to the best of the authors' knowledge. In terms of data and patient distribution, there are a total of 5,210 images from 2,815 SARS-CoV-2 positive patients and 13,993 images from 13,851 SARS-CoV-2 negative patients. The negative patient cases comprise of both no pneumonia and non-SARS-CoV-2 pneumonia patient cases, with 8,418 no pneumonia images from 8,300 patients and 5,575 non-SARS-CoV-2 pneumonia images from 5,551 patients. The distribution of CXR images in the benchmark dataset for SARS-CoV-2 negative and positive cases is shown in Figure~\ref{fig:image-dist}, with respective patient distribution shown in Figure~\ref{fig:patient-dist}.  Select patient cases from the benchmark dataset were reviewed and reported on by two board-certified radiologists with 10 and 19 years of experience, respectively.

The COVID-Net CXR-2 network is evaluated on a balanced test set of 200 SARS-CoV-2 positive images from 178 patients and 200 SARS-CoV-2 negative images from 100 no pneumonia and 100 non-SARS-CoV-2 pneumonia patient cases. The test images were randomly selected from international patient cohorts curated by the Radiological Society of North America (RSNA)~\cite{kaggle,ricord}, with the cohorts collected and expertly annotated by an international group of scientists and radiologists from different institutes around the world.  The test set was selected in such a way to ensure no patient overlap between training and test sets.

Table~\ref{tab:demographic} summarizes the demographic variables and imaging protocol variables of the CXR data in the benchmark dataset. It can be observed that the patient cases in the cohort used in the benchmark dataset are distributed across the different age groups, with the mean age being 46.89 and the highest number of patients in the cohort being between the ages of 50-59.  

The benchmark dataset, along with all data generation and preparation scripts, are available in an open source manner at \url{http://www.covid-net.ml}.

\begin{table}[h] 
\centering
\caption{Summary of demographic variables and imaging protocol variables of CXR data in the benchmark dataset. Age and sex statistics are expressed on a patient level, while imaging view statistics are expressed on an image level.}
\label{tab:demographic}
\begin{tabular}{|c|c|c|}
\hline
\multicolumn{1}{|l|}{\textbf{Age}} & mean $\pm$ std & $46.89 \pm 17.65$ \\ \hline
 & \textless 20 & 1026 (6.2\%) \\ \hline
 & 20-29 & 1821 (10.9\%) \\ \hline
 & 30-39 & 2268 (13.6\%) \\ \hline
 & 40-49 & 2908 (17.5\%) \\ \hline
 & 50-59 & 3486 (20.9\%) \\ \hline
 & 60-69 & 2358 (14.2\%) \\ \hline
 & 70-79 & 1010 (6.1\%) \\ \hline
 & 80-89 & 300 (1.8\%) \\ \hline
 & 90+ & 86 (0.5\%) \\ \hline
 & Unknown & 1393 (8.4\%) \\ \hline
\multicolumn{3}{|l|}{\textbf{Sex}} \\ \hline
 & Male & 8774 (52.7\%) \\ \hline
 & Female & 6768 (40.6\%) \\ \hline
 & Unknown & 1114 (6.7\%) \\ \hline
\multicolumn{3}{|l|}{\textbf{Imaging view}} \\ \hline
 & PA & 9321 (48.5\%) \\ \hline
 & AP & 7307 (38.1\%) \\ \hline
 & Unknown & 2575 (13.4\%) \\ \hline
\end{tabular}%
\end{table}

\subsection{Network design and learning}
Leveraging the aforementioned benchmark dataset, we built COVID-Net CXR-2 to be tailored for COVID-19 case detection from CXR images using machine-driven design. The machine-driven design exploration strategy automatically discovers highly customized and unique macroarchitecture and microarchitecture designs to optimize the trade-off between accuracy and efficiency.  More specifically, the concept of generative synthesis~\cite{gensynth} was leveraged, where the macroarchitecture and microarchitecture designs of a tailored deep neural network architecture are determined by an optimal generator $\mathcal{G}$ whose generated deep neural network architectures $\left\{N_s|s \in S\right\}$ maximize a universal performance function $\mathcal{U}$ (e.g., \cite{wong2018netscore}), with constraints imposed on a set of operational requirements as defined by an indicator function $1_r(\cdot)$,

\begin{equation}
\mathcal{G}  = \max_{\mathcal{G}}~\mathcal{U}(\mathcal{G}(s))~~\textrm{subject~to}~~1_r(\mathcal{G}(s))=1,~~\forall s \in S.
\label{eqoptimization}
\end{equation}

\noindent where $S$ denotes a set of seeds to the generator.  For the purpose of building COVID-Net CXR-2, the set of constraints imposed via indicator function $1_r(\cdot)$ were: (1) sensitivity $\geq$ 95\%, and (2) positive predictive value (PPV) $\geq$ 95\%. 

\begin{figure*}[t]
  \centering
  \includegraphics[width= \linewidth]{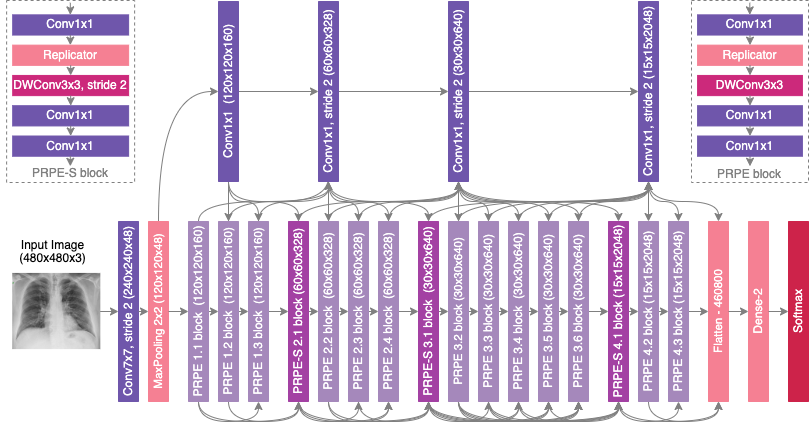}
  \caption{The proposed COVID-Net CXR-2 architecture design.  The COVID-Net design exhibits high architectural diversity and sparse long-range connectivity, with macro and microarchitecture designs tailored specifically for the detection of COVID-19 from chest X-ray images. The network design leverages light-weight design patterns in the form of projection-replication-projection-expansion (PRPE) patterns to provide enhanced representational capabilities while maintaining low architectural and computational complexities.}
  \label{fig:architecture}
\end{figure*}

A number of observations can be made about the proposed COVID-Net CXR-2 deep convolutional neural network architecture design shown in Figure~\ref{fig:architecture}.  It can be observed that the proposed COVID-Net CXR-2 deep convolutional neural network possesses a light-weight network architecture design that exhibits notable diversity from both a macroarchitecture and microarchitecture design perspective.  More specifically, the COVID-Net CXR-2 architecture design possesses a diverse mix of point-wise and depth-wise convolutions, and a very sparing use of conventional convolutions at the input stage of the architecture.  In particular, the network design leverages light-weight design patterns in the form of project-replication-projection-expansion (PRPE) patterns to provide enhanced representational capabilities while maintaining low architectural and computational complexities.  Furthermore, sparse use of long-range connections can also be observed in the network architecture design to strike a good balance between architectural and computational efficiency and representational capacity.  The strong balance between efficiency and accuracy achieved by the proposed network highlights the utility of machine-driven design exploration for tailored architectures beyond the capabilities of manual, human designs. 

Training was conducted using a binary cross-entropy loss and Adam optimization with learning rate of 1e-5 on a batch size of 8 for 40 epochs. The final model was selected by tracking the validation accuracy throughout training and employing early stopping. All construction, training, and evaluation are conducted in the TensorFlow deep learning framework. As a pre-processing step, the CXR images were cropped (top 8\% of the image) prior to training and testing to better mitigate commonly-found embedded textual information. The CXR images were then resampled to 480$\times$480 and normalized to the range [0, 1] via division by 255. Furthermore, data augmentation was leveraged during training with the following augmentation types: translation ($\pm$ 10\% in x and y directions), rotation ($\pm$ 10$^{\circ}$), horizontal flip, zoom ($\pm$ 15\%), and intensity shift ($\pm$ 10\%). A batch re-balancing strategy was introduced to promote better distribution of SARS-CoV-2 positive cases and SARS-CoV-2 negative cases at a batch level.

The COVID-Net CXR-2 network and associated scripts are available in an open source manner at \url{http://www.covid-net.ml}.

\subsection{Explainability-driven performance validation}
The trained COVID-Net CXR-2 network was audited to gain deeper insights into its decision-making behaviour and ensure that it is driven by clinically relevant indicators rather than erroneous cues such as imaging artifacts and embedded metadata. We leveraged GSInquire~\cite{gsinquire} to conduct explainability-driven performance validation, as it was shown to provide state-of-the-art explanations compared to other methods in literature. More specifically, GSInquire takes advantage of the generative synthesis~\cite{gensynth} strategy leveraged during machine-driven design exploration to identify and visualize the critical factors that COVID-Net CXR-2 uses to make predictions through an inquisitor $\mathcal{I}$ within a generator-inquisitor pair $\left\{\mathcal{G},\mathcal{I}\right\}$, where the generator $\mathcal{G}$ is the optimal generator used to generate COVID-Net CXR-2 as shown in Equation~\ref{eqoptimization}.  Compared to other explainability methods in literature that produce relative heat maps that visualize variations in potential importance within an image, GSInquire has a unique capability of surfacing specific critical factors within an image that quantitatively impact the decisions made by the deep neural network.  This makes the explanations easier to interpret objectively and better reflects the decision-making process of the deep neural network for validation purposes.

Explainability-driven performance validation is crucial for improved transparency and trust, particularly in healthcare applications such as clinical decision support. It can also help clinicians to uncover new insights into key visual indicators associated with COVID-19 to improve screening accuracy. 

\subsection{Radiologist validation}
To further audit the results for COVID-Net CXR-2, select patient cases from the explainability-driven performance validation were further reviewed and reported on by two board-certified
radiologists (A.S. and A.A.). The first radiologist (A.S.) has over 10 years of experience, while the
second radiologist (A.A.) has over 19 years of radiology experience.

\section{Results and Discussion}\label{sec:results}

To explore and evaluate the efficacy of the proposed COVID-Net CXR-2 deep convolutional neural network design for detecting COVID-19 cases from CXR images, we conducted a quantitative performance analysis to assess its architectural and computational complexity as well as its detection performance on the benchmark dataset.  We further explored its decision-making behaviour using an explainability-driven performance validation approach to audit COVID-Net CXR-2 in a transparent and responsible manner.  The quantitative and qualitative results are presented and discussed in detail below.

\subsection{Quantitative analysis}
Let us first explore the quantitative performance and underlying complexity of the proposed COVID-Net CXR-2 deep neural network architecture tailored for the detection of COVID-19 cases from CXR images.  For comparison purposes, we also provide quantitative results on the test data for COVID-Net~\cite{covidnet}, which was shown to provide state-of-the-art performance for COVID-19 detection when compared to other methods in research literature, and ResNet-50~\cite{resnetv2}, another state-of-the-art deep neural network commonly leveraged in computer vision. The ResNet-50 referenced in this study was trained using the same proposed CXR benchmark dataset and augmentation practices, with the binary cross-entropy loss and Adam optimization tuned to a learning rate of 1e-2 and batch size of 16 for 40 epochs for optimal performance. The architectural and computational complexity of COVID-Net CXR-2 in comparison to COVID-Net~\cite{covidnet} and ResNet-50~\cite{resnetv2} is shown in Table~\ref{tab:quant_results1}, with quantitative performance results shown in Table~\ref{tab:quant_results2}. It can be observed that the COVID-Net CXR-2 network achieves 5.8\% and 2.3\% higher test accuracy (at 96.3\%) than both state-of-the-art ResNet-50 and COVID-Net networks respectively, while maintaining the lowest network complexity. For instance, the COVID-Net CXR-2 has $\sim$65\% and $\sim$25\% lower architectural complexity (at $\sim$8.8M parameters), and $\sim$68\% and $\sim$25\% lower computational complexity (at $\sim$5.6G MACs) than both networks respectively. It can also be observed that the COVID-Net CXR-2 network was able to outperform both networks in regards to sensitivity, with 7\% and 2\% higher sensitivity when compared to both networks respectively (at 95.5\%), and achieved 6.5\% higher positive predictive value (PPV) than the ResNet-50 and 3\% lower PPV than the COVID-Net network (at 97.0\%). This trade-off of higher sensitivity gained by COVID-Net CXR-2 compared to COVID-Net in exchange for a decrease in PPV (which is still quite high for COVID-Net CXR-2 at 97.0\%) is a reasonable one given that a higher sensitivity results in fewer missed SARS-CoV-2 positive patient cases during the screening process. This is very important from a clinical perspective in controlling the spread of the SARS-CoV-2 virus during the on-going COVID-19 pandemic in light of the new highly infectious variants. Finally, Table~\ref{tab:conf_matrix} provides a more detailed picture of the performance of COVID-Net CXR-2 via the confusion matrix.

\subsection{Qualitative analysis}

Examples of patient cases and the associated critical factors identified by GSInquire as the driving factors behind the decision-making behaviour of COVID-Net CXR-2 are shown in Figure~\ref{fig:explainability}. It can be observed that the network primarily leverages areas in the lungs of the CXR images and is not relying on incorrect factors such as artifacts outside of the body, motion artifacts, and embedded markup symbols. From further investigation into the correctly detected COVID-19 cases, the critical factors typically identified correspond to clinically relevant visual cues such as ground-glass opacities, bilateral abnormalities, and interstitial abnormalities. These observations indicate that the network's decision-making process is generally consistent with clinical interpretation.  

\begin{table}[!t]
\caption{Architectural and computational complexity of COVID-Net CXR-2 network in comparison to COVID-Net~\cite{covidnet} and ResNet-50~\cite{resnetv2} architectures. Best results highlighted in \textbf{bold}. }
\begin{center}
\begin{tabular}{|c|c|c|}
\hline
     \textbf{Architecture} & \textbf{Parameters (M)} & \textbf{MACs (G)} \\
\hline
ResNet-50~\cite{resnetv2} & 24.97 & 17.75\\
\hline
COVID-Net~\cite{covidnet} & 11.8 & 7.5\\
\hline
 COVID-Net CXR-2 & \textbf{8.8} & \textbf{5.6}\\
\hline
\end{tabular}\par
\label{tab:quant_results1}
\end{center}
\end{table}

\begin{table}[!t]
\caption{Sensitivity, positive predictive value (PPV), and accuracy of COVID-Net CXR-2 on the test data from the CXR benchmark dataset in comparison to COVID-Net ~\cite{covidnet} and ResNet-50~\cite{resnetv2} architecture.  Best results highlighted in \textbf{bold}.}
\begin{center}
\begin{tabular}{|c|c|c|c|}
\hline
     \textbf{Architecture} & \textbf{Sensitivity ($\%$)} & \textbf{PPV ($\%$)} & \textbf{Accuracy ($\%$)} \\
\hline
ResNet-50~\cite{resnetv2} & 88.5 & 92.2 & 90.5\\
\hline
 COVID-Net~\cite{covidnet} & 93.5 & \textbf{100} & 94.0 \\ 
\hline
 COVID-Net CXR-2 & \textbf{95.5} & 97.0 &  \textbf{96.3} \\
\hline
\end{tabular}\par
\label{tab:quant_results2}
\end{center}
\end{table}

\begin{table}[!t]
\caption{Confusion matrix of COVID-Net CXR-2 network. }
\begin{center}
\begin{tabular}{|c|c|c|}
\hline
     \textbf{SARS-CoV-2} & \textbf{Negative} & \textbf{Positive} \\
\hline
\textbf{Negative} & 194 & 6\\
\hline
\textbf{Positive} & 9 & 191\\
\hline
\end{tabular}\par
\label{tab:conf_matrix}
\end{center}
\end{table}

This explainability-driven performance validation process is important for a number of important reasons from the perspectives of transparency, dependability, and trust. First of all, this process enabled us to audit and validate that COVID-Net CXR-2 exhibits dependable decision-making behaviour since it is not only guided by clinically relevant visual indicators, but more importantly it is not dependent on erroneous visual indicators such as imaging artifacts, embedded markup symbols, and embedded text in the CXR images. This ensures the network does not make the right decisions for the wrong reasons.  Second, this validation process allows for the discovery and identification of potential new insights into what types of clinically relevant visual indicators are particularly useful for differentiating between SARS-CoV-2 infections and non-SARS-CoV-2 infections.  Such discoveries could be useful information for aiding clinicians and radiologists in better detecting SARS-CoV-2 infection cases during the clinical decision process.  Finally, by validating the behaviour of COVID-Net CXR-2 in a transparent and responsible manner, one can provide greater transparency and garner greater trust for clinicians and radiologists during usage in their screening process to make faster yet accurate assessments.

These quantitative and qualitative results show that COVID-Net CXR-2 not only provides strong COVID-19 detection performance, but also exhibits clinically relevant decision-making behaviour.  

\subsection{Radiologist analysis}
\begin{figure*}[!t]
  \centering
  \includegraphics[width= \linewidth]{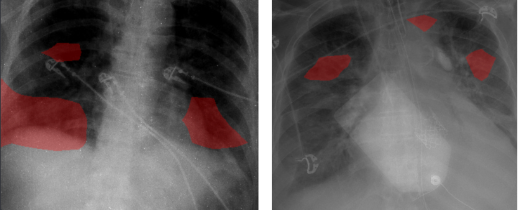}
  \caption{Examples of patient cases and the associated critical factors (highlighted in red) as identified by GSInquire~\cite{gsinquire} during explainability-driven performance validation as what drove the decision-making behaviour of COVID-Net CXR-2. (left) Case 1, (right) Case 2.  Radiologist validation showed that several of the critical factors identified are consistent with radiologist interpretation.}
  \label{fig:explainability}
\end{figure*}

The expert radiologist findings and observations with regards to the critical factors identified by GSInquire for select patient cases shown in Figure~\ref{fig:explainability} are as follows. In both cases, COVID-Net CXR-2 detected them to be patients with SARS-CoV-2 infection, which were clinically confirmed.

\textbf{Case 1.} According to radiologist findings, it was observed by both radiologists that there is an opacity at the right lung base, which is consistent with one of the identified critical factors leveraged by COVID-Net CXR-2. Additional imaging would be recommended by both radiologists.

\textbf{Case 2.} According to radiologist findings, it was observed by both radiologists that there are opacities in the right midlung and left paratracheal region that coincide with the identified critical factors leveraged by COVID-Net CXR-2 in that region.  Additional imaging would be recommended by one of the radiologists.

As such, based on the radiologist findings and observations on the two patient cases, it was demonstrated that several of the identified critical factors leveraged by COVID-Net CXR-2 are consistent with radiologist interpretation.

\section{Conclusion}\label{sec:conclusion}
In this study, we introduced COVID-Net CXR-2, an enhanced deep convolutional neural network design tailored for COVID-19 detection from CXR images that is built based on a greater quantity and diversity of patient cases than the original COVID-Net.  A new benchmark dataset of CXR images representing a multinational cohort of 16,656 patients from at least 51 countries was also introduced, which is the largest, most diverse COVID-19 CXR dataset in open access form to the best of the authors' knowledge.  Experimental results demonstrate that COVID-Net CXR-2 can not only achieve strong COVID-19 detection performance in terms of accuracy, sensitivity, and PPV, but also exhibit behaviour consistent with clinical interpretation during an explainability-driven performance validation process, which was further validated based on radiologist interpretation.  The hope is that the release of COVID-Net CXR-2 and its respective benchmark dataset in an open source manner will help encourage researchers, clinical scientists, and citizen scientists to accelerate advancements and innovations in the fight against the pandemic.  Further work involves the continued improvement of the benchmark dataset as well as architecture design, as well as exploration into other clinical workflow tasks (e.g. severity assessment, treatment planning, resource allocation, etc.) as well as other imaging modalities (e.g. computed tomography, point-of-care ultrasound, etc.).  Furthermore, we aim to conduct more comprehensive auditing of both the benchmark dataset as well as the deep neural network.
    
\begin{ack}
    We thank the Natural Sciences and Engineering Research Council of Canada (NSERC), the Canada Research Chairs program, the Canadian Institute for Advanced Research (CIFAR), DarwinAI Corp., and the organizations and initiatives from around the world collecting valuable COVID-19 data to advance science and knowledge. The study has received ethics clearance from the University of Waterloo (42235).
\end{ack}

\section*{Author contributions statement}
    M.P., N. T., A.C., and A.W. conceived the experiments, M.P., N. T., A.Z., S.S., H.A., and H.G. conducted the experiments, all authors analysed the results, A.A. and A.S. reviewed and reported on select patient cases and corresponding explainability results illustrating model's decision-making behaviour, and all authors reviewed the manuscript.
    
\section*{Declaration of interests}
A.W. and A.C. are affiliated with DarwinAI Corp.

\printbibliography

\end{document}